\input{aipcheck}

\documentclass[numbers,final]{aipproc}

\layoutstyle{8x11single}

\begin{document}

\title{Broadening in Medium-induced QCD  Radiation off a $q\bar q$ Antenna}



\author{N\'estor Armesto}{
  address={Departamento de F\'isica de Part\'iculas and IGFAE, 
Universidade de Santiago de Compostela, 
E-15782 Santiago de Compostela, 
Galicia-Spain}
}

\author{Hao Ma}{
  address={Departamento de F\'isica de Part\'iculas and IGFAE, 
Universidade de Santiago de Compostela, 
E-15782 Santiago de Compostela, 
Galicia-Spain}
}

\author{Yacine Mehtar-Tani}{
  address={Departamento de F\'isica de Part\'iculas and IGFAE, 
Universidade de Santiago de Compostela, 
E-15782 Santiago de Compostela, 
Galicia-Spain}
}

\author{Carlos A. Salgado}{
  address={Departamento de F\'isica de Part\'iculas and IGFAE, 
Universidade de Santiago de Compostela, 
E-15782 Santiago de Compostela, 
Galicia-Spain}
}

\author{Konrad Tywoniuk}{
  address={Department of Astronomy and Theoretical Physics, 
Lund University, S\"olvegatan 14A,
S-223 62 Lund, 
Sweden}
}

\begin{abstract}
In this contribution, the one-gluon radiation spectrum off a massive quark-antiquark ($q {\bar q}$) antenna traversing a QCD medium is calculated in the eikonal approximation. The interference between emissions from the quark and the antiquark is considered. The gluon spectrum computed at first order in the opacity expansion is collinear finite but infrared divergent, which is in contrast with the result obtained from an independent emitter which is both infrared and collinear finite. Phenomenological consequences on the broadening of the emitted gluon are investigated. In the soft gluon emission limit, the interference between emitters causes a different broadening style from the typical {\bf k}-broadening in the case of an independent emitter.
\end{abstract}

\maketitle

\section{Introduction}
Based on  perturbative QCD, in the soft and collinear limits the single-inclusive gluon spectrum was calculated in \cite{BDMPS1, BDMPS2, BDMPS3, BDMPS4, BDMPS5, BDMPS6} by re-summing arbitrary orders of the background scattering centers, and in \cite{BDMPS6,GLV1, GLV2} by calculating the background fields order by order - the so-called opacity expansion. The spectrum of these two equivalent approaches, denoted as BDMPS-Z-W/GLV, is both infrared and collinear finite. This is because the dominant contribution is from the rescattering of the emitted off-shell gluon with the medium. The gluon energy distribution shows the typical ${\bf k}$-broadening \citep{Baier:2001qw,Salgado:2003rv}, and one has $\Delta E$ $\propto$ $\langle {\bf k}^2\rangle$. The mentioned calculations, however, do not contain the effects of the interference between different emitters.

The interference effect was first studied in vacuum, i.e., a quark-antiquark ($q {\bar q}$) antenna was considered, see \cite{book} and references therein. The soft gluon radiation spectrum off a massless $q {\bar q}$ antenna in vacuum exhibits angular ordering (AO): radiation is suppressed at $\theta$ $>$ $\theta_{q {\bar q}}$ (after averaging over the gluon azimuthal angle), where $\theta$ is the angle between the emitted gluon and the parent quark (antiquark) and $\theta_{q {\bar q}}$ is the antenna opening angle. The massless antenna spectrum diverges when $\theta$ $\rightarrow$ $0$ and $\omega$ $\rightarrow$ $0$. For the heavy quarks, on the other hand, AO is modified and the collinear divergence disappears due to the dead cone effect, but the soft divergence remains. The result was shown to hold for arbitrary color representations of the $q\bar q$ pair. Medium-induced soft gluon radiation off a massless $q {\bar q}$ antenna has been studied recently \cite{MSTprl, MSTdecoh, MT, CI}. The spectrum exhibits antiangular ordering (AAO) - there is no collinear divergence ($\theta$ $>$ $\theta_{q {\bar q}}$ $>$ $0$) - but the soft divergence still exists, at variance with the results in BDMPS-Z-W/GLV. Again, the result is valid not only for color singlet states but also for arbitrary color representations. Mass effect and coherence are studied in \cite{QM2011}, where we find that more collimated antennas lose less energy and the size of the mass effect of the antenna is similar to that of the independent emitter. In this work we continue to analyze the medium-induced QCD radiation off a massive antenna, for a static medium at first order in opacity (one scattering). Due to space limitations, here we show some selected results on the broadening of the emitted gluon.

\section{Results}
\begin{figure}[t!]
\includegraphics[width=0.8\textwidth]{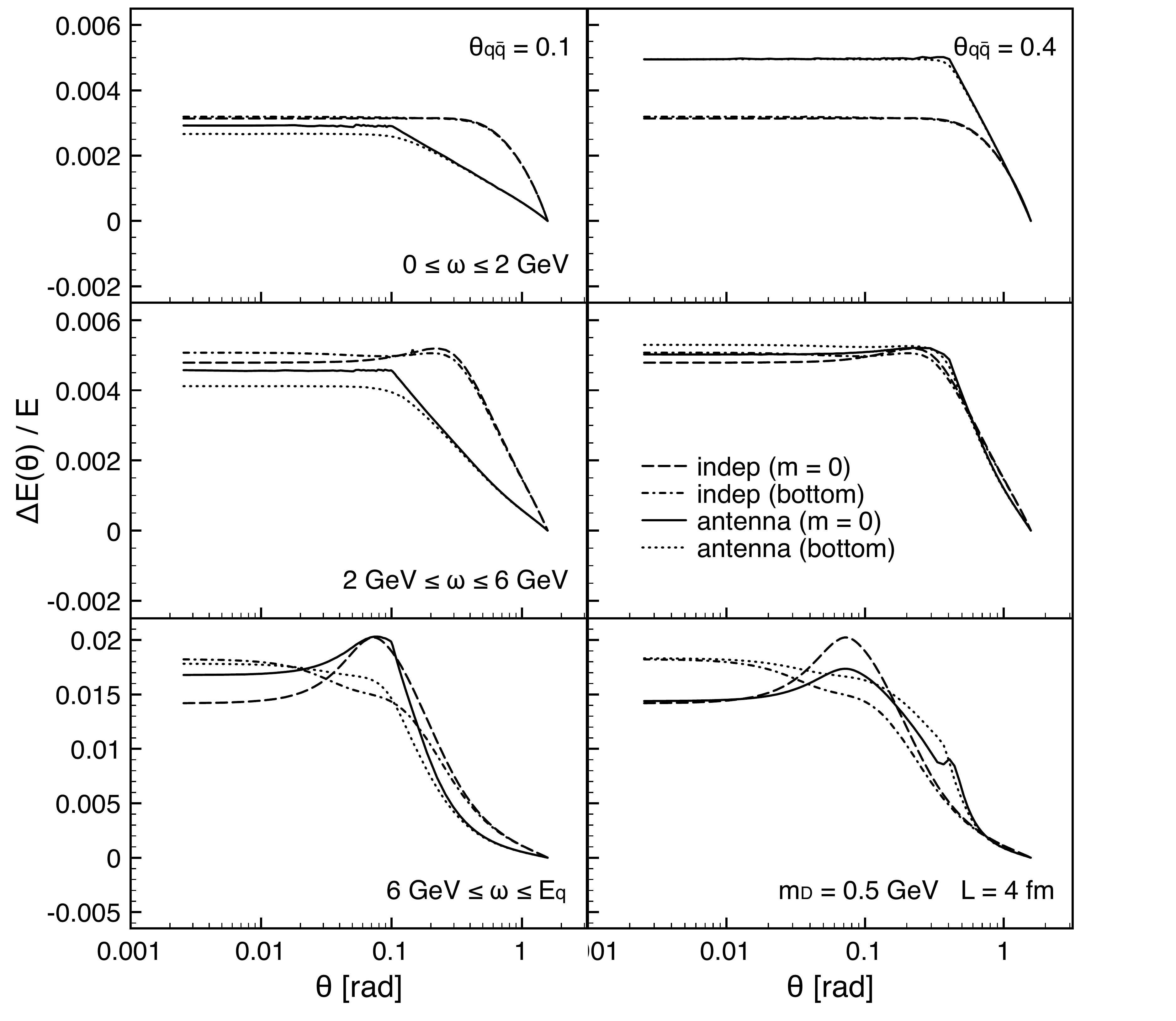}
\caption{\label{HELT}Dependence of the medium-induced radiative relative energy loss outside a cone on the angle defining the cone. The parameters are: Debye mass $m_D=0.5$ GeV, medium length $L=4$ fm, and $\theta_{q\bar q}=0.1(0.4)$ for the plots on the left (right). The solid curves correspond to the massless antenna, the dotted curves to the bottom antenna, the dashed curves to the massless independent spectra and the dash-dotted curves to the bottom independent spectra. From top to bottom, the values used in Eq. (\ref{deoep}) for $\omega_{\rm min}$ are 0, 2 and 6 GeV, while those for $\omega_{\rm max}$ are 2, 6 and $E_q$.}
\end{figure}

In order to investigate the relation between the average energy loss and the broadening of the antenna and to compare it with the one of the independent emitter, the medium-induced radiative energy loss outside of a given emission angle $\theta$ is computed via
\begin{center}
\begin{equation}
\Delta E (\theta) = \int_{\omega_{\rm min}}^{\omega_{\rm max}} {\rm d} \omega \int_{\theta}^{\pi / 2} {\rm d} \theta' \, \omega \frac{{\rm d} N}{{\rm d} \omega \, {\rm d} \theta'}\ .
\label{deoep}
\end{equation}
\end{center}
The ratio $\Delta E (\theta) / E$ as a function of $\theta$ is shown in Fig. \ref{HELT}, where $E=E_q=100$ GeV for the independent emitter case and for the antenna the spectrum is divided by 2 as it obtains contributions from both the quark and antiquark. Debye mass $m_D=0.5$ GeV and medium length $L=4$ fm are used (Fig. \ref{HELT}). Note that the behavior of $\Delta E (\theta)$ with the angle $\theta$ traces the angular behavior of the differential energy spectrum: A decrease slower than linear of $\Delta E (\theta)$ comes from an increasing energy spectrum, a decreasing linear  behavior of $\Delta E (\theta)$ results from a flat energy spectrum, and a decrease stronger than linear of $\Delta E (\theta)$ traces a falling energy spectrum\footnote{A flat behavior of $\Delta E (\theta)$ indicates the null contribution, which is due to the antiangular ordering for the antenna. An increasing behavior of $\Delta E (\theta)$ indicates the existence of negative contributions at small angles due to destructive interference, a well-known phenomenon in the BDMPS-Z-W/GLV framework \citep{ASW} for small medium parameters.}. The first behavior points to ${\bf k}$-broadening of the radiation which reaches the upper bound at a finite angle.

By the examination of Fig. \ref{HELT}, we see that the independent emitter case  exhibits, in the regions of small and moderate $\omega$ (i.e., $0$ $\leq$ $\omega$ $\leq$ $2$ GeV and $2$ GeV $\leq$ $\omega$ $\leq$ $6$ GeV), ${\bf k}$-broadening which emerges due to the rescattering of the emitted off-shell gluon with the medium as discussed in the Introduction. The dominant contribution of the antenna, in the limit $\omega\rightarrow 0$, only contains the on-shell gluon bremsstrahlung and the rescatterings of the hard quark and antiquark with the medium, so there is no ${\bf k}$-broadening for the antenna. For $\theta_{q {\bar q}}=0.1$ in the regions of small and moderate $\omega$, $\Delta E (\theta)$ of the massless antenna is almost a constant for $\theta\leq\theta_{q {\bar q}}$ due to antiangular ordering. For $\theta>\theta_{q {\bar q}}$, the curve of the massless antenna drops monotonously and faster than linear with  increasing gluon emission angle. The suppression of the gluon radiation off the bottom antenna (dotted curve) as compared with the one off the massless antenna (solid curve) can be clearly seen at $\theta\leq\theta_{q {\bar q}}$, because most of the gluons are emitted around the opening angle. For  $\theta_{q {\bar q}}=0.4$, the antenna still keeps the interference feature - flat behavior for $\theta<\theta_{q\bar q}$ - in the soft gluon emission sector, but it shows some ${\bf k}$-broadening in the moderate gluon emission sector. Note that in the region of small $\omega$ (see the top panels in Fig. \ref{HELT}) and inside the antenna opening angle, the dead cone effect is not clear for the antenna of $\theta_{q {\bar q}}=0.4$ as compared with the one of $\theta_{q {\bar q}}=0.1$. The interference between emitters included in the antenna generates more gluon radiation at $\theta_{q {\bar q}}=0.4$ than at $\theta_{q {\bar q}}=0.1$ for the specific choice of the parameters, i.e., $m_D=0.5$ GeV, $L=4$ fm and $0$ $\leq$ $\omega$ $\leq$ $2$ GeV. In the region of large $\omega$ (i.e., $6$ GeV $\leq$ $\omega$ $\leq$ $E_q$), for both opening angles $\theta_{q {\bar q}}=0.1$ and $0.4$,  the antenna and the independent emitter spectrum exhibit similar features with respect to broadening in both the massless and the massive cases. There is a second peak at around $\theta = 0.4 \, (= \theta_{q {\bar q}})$ for the massless antenna in the bottom right panel in Fig. \ref{HELT}, which is the peak of the antiquark spectrum inside the antenna and shows the decoherence of the $q {\bar q}$ pair at large antenna opening angle and large gluon energy. One can see that the interference spectrum dominates  gluon radiation when the antenna opening angle is small and the emitted gluon is soft, and then the antenna exhibits the new broadening style - antiangular ordering; while the antenna behaves like a superposition of independent emitters when the opening angle is large and the radiated gluon is hard, and then the antenna shows the typical ${\bf k}$-broadening.

\section{Conclusions}
In this contribution, some results of the medium-induced gluon radiation spectrum off a $q {\bar q}$ antenna at first order in opacity for the massive case are shown. The average energy loss outside of a given emission angle $\theta$ is studied, and we find that there is no typical ${\bf k}$-broadening in the antenna in either the massless or the massive cases for small opening angles of the antenna and small energies of the emitted gluon. In the soft gluon emission region, the dead cone effect becomes more and more important as the antenna opening angle decreases. The interference between emitters included in the antenna opens the phase space for soft gluon radiation at relatively large opening angles for some specific choices of the parameters. We find that the antenna radiation is dominated by the case of independent emitter for large opening angles of the antenna and for large energies of the emitted gluon.


\begin{thebibliography}{9}

\bibitem{BDMPS1}
R. Baier, Yu. L. Dokshitzer, A. H. Mueller, S. Peign\'e, and D. Schiff, Nucl. Phys. B {\bf 483}, 291 (1997).

\bibitem{BDMPS2}
R. Baier, Yu. L. Dokshitzer, A. H. Mueller, S. Peign\'e, and D. Schiff, Nucl. Phys. B {\bf 484}, 265 (1997).

\bibitem{BDMPS3}
B. G. Zakharov, JETP Lett. {\bf 63}, 952 (1996).

\bibitem{BDMPS4}
B. G. Zakharov, JETP Lett. {\bf 65}, 615 (1997).

\bibitem{BDMPS5}
U. A. Wiedemann, Nucl. Phys. B {\bf 582}, 409 (2000).

\bibitem{BDMPS6}
U. A. Wiedemann, Nucl. Phys. B {\bf 588}, 303 (2000).

\bibitem{GLV1}
M. Gyulassy, P. Levai, and I. Vitev, Phys. Rev. Lett. {\bf 85}, 5535 (2000).

\bibitem{GLV2}
M. Gyulassy, P. Levai, and I. Vitev, Nucl. Phys. B {\bf 594}, 371 (2001).

\bibitem{Baier:2001qw}
R.~Baier, Y.~L.~Dokshitzer, A.~H.~Mueller, and D.~Schiff, Phys.\ Rev.\  {\bf C64}, 057902 (2001).

\bibitem{Salgado:2003rv}
C.~A.~Salgado, U.~A.~Wiedemann, Phys.\ Rev.\ Lett.\  {\bf 93}, 042301 (2004).

\bibitem{book}
Yu. L. Dokshitzer, V. A. Khoze, A. H. Mueller, and S. I. Troyan, {\it ``Basics of Perturbative QCD'', Gif-sur-Yvette, France, Ed. Fronti\`eres 1991}.

\bibitem{MSTprl}
Y. Mehtar-Tani, C. A. Salgado, and K. Tywoniuk, Phys. Rev. Lett. {\bf 106}, 122002 (2011).

\bibitem{MSTdecoh}
Y. Mehtar-Tani, C. A. Salgado, and K. Tywoniuk, arXiv:1102.4317v1 [hep-ph].

\bibitem{MT}
Y. Mehtar-Tani and K. Tywoniuk, arXiv:1105.1346v1 [hep-ph].

\bibitem{CI}
J. Casalderrey-Solana and E. Iancu, arXiv:1105.1760v2 [hep-ph].

\bibitem{QM2011}
N. Armesto, H. Ma, Y. Mehtar-Tani, C. A. Salgado and K. Tywoniuk, arXiv:1107.0291v1 [hep-ph].

\bibitem{ASW}
N. Armesto, C. A. Salgado, and U. A. Wiedemann, Phys. Rev. D {\bf 69}, 114003 (2004).

\end{thebibliography}
\end{document}